\documentclass{article}
\usepackage{graphicx}
\usepackage{dcolumn}
\usepackage{bm}
\usepackage{pstricks}
\usepackage{epsfig}
\begin{document}
\newtheorem{thm}{Theorem}
\newtheorem{cor}{Corollary}
\newtheorem{Def}{Definition}
\newtheorem{lem}{Lemma}
\begin{center}
{\large \bf SU(2) Relativity and the EPR Paradox} \vspace{5 mm}

Paul O'Hara
\\
\vspace{5mm} {\small\it
Dept. of Mathematics\\
\baselineskip 12pt
Northeastern Illinois University\\
5500 North St. Louis Avenue\\
Chicago, IL 60625-4699, USA.\\
\vspace{5mm} email: pohara@neiu.edu \\}
\end{center}
\vspace{10mm}

\begin{abstract} In the normal presentation of the EPR
problem a comparison is made between the (weak) Copenhagen
interpretation of quantum mechanics which seems to suggest that at
times action at a distance may take place, and the hidden parameter
interpretation which must satisfy Bell's inequality, in
contradiction to the predictions of quantum mechanics. In this
paper, we consider a relativistic approach to the paradox. However,
the frame of reference under consideration is not the usual Lorenz
frame but rather the spin frame of reference which is invariant with
respect to the SU(2) group. \newline Key WORDS: hidden variables,
Copenhagen, SU(2) relativity
\end{abstract}

\section {INTRODUCTION}

Before giving an alternative approach to the EPR problem, we must
first be clear as to what the paradox is and what a proper solution
should entail: Consider two particles emitted in the singlet state.
Let $\lambda_1$ be the measurement of spin made on particle 1 in the
direction $\vec n_1$. By definition of a singlet state, a
measurement of the spin of particle 2 will yield the value
$—\lambda_1$, if measured in the same direction $\vec n_1$, a fact
also observed in experiment.

The paradox arises when two competing epistemological views of
reality are brought to bear on the matter. In the naive realist
interpretation \cite{heal} (as it has come to be called) the
particles have a definite predetermined spin value. From this
perspective, reality can be known by simply ``looking'' at it\cite
{lon}. What we see is really what is there, and moreover, based on
these assumptions, a naive realist can construct a mathematical
model of spin which gives rise either to Bell's inequality
\cite{Bell} or the GHZ theorem \cite{green}. In either case, a
mathematical contradiction arises which cannot be explained by naive
realism.

The weak Copenhagen interpretation of QM, on the other hand, also
runs into a difficulty. At the core of this interpretation is the
projection postulate which effectively says that the act of
measurement forces the particle to choose between the two
alternatives. However, if this is the case, it means that the
particles in the singlet state choose equal but opposite values on
measurement, although communication between the particles has been
eliminated. This state of affairs is normally explained by saying
that the interaction is ``non-local'' which is equivalent to saying
that there is action at a distance, and some perhaps would interpret
this to mean that cause and effect have broken down.

Therein lies the paradox. Naive realism gives rise to a
contradiction while the projection postulate, if taken literally
gives rise to action at a distance, which appears to be a violation
of relativity theory.

The epistemological approach taken here is to view the spin state as
preexisting while at the same time considering the measured value of
the spin as an SU(2) relativity effect correlated to the direction
of measurement, a fact previously noted in other publications \cite{
LAT, NYAS}. In other words, the same spin state can be interpreted
as +1 or -1 depending on our point of view. Consequently, as we
shall see below, this permits two different values to be assigned to
the same event, and will result in Bell's inequality, if the SU(2)
relativity effect is ignored.

An analogy might help. The Earth as viewed from the North pole can
be seen as rotating in an anti-clockwise direction (-1), while the
same state of motion can be viewed as being in a clockwise direction
(+1), if viewed from the South pole. However, in the notation of
quantum mechanics this same event, as seen from the equator, cannot
be interpreted in a consistent way and it is this inconsistency that
gives rise to both Bell's inequality and the GHZ result. In other
words, the same state can take on two different values, depending on
the viewpoint and since the essence of Bell's inequality involves
correlations from three different directions, its a prime candidate
for bringing forth inconsistencies, unless this SU(2) relativity
effect is taken into account \cite{LAT}. Another useful analogy is
to envisage a Mobius strip and ask what is the orientation of the
surface? It has no orientation, although from any one perspective it
can be considered a ``two-faced''surface. Once again, orientation
depends on the point of view \cite{NYAS}.

\section {DIFFERENT REPRESENTATIONS} The first thing to grasp is that
there are two equal but different representations of spin.
Specifically, consider the following:
\begin{equation}
    \left(
            \begin{array}{cc}
              1 & 0 \\
              0 &-1 \\
            \end{array}
          \right)
 \left(
 \begin{array}{c}
    1 \\
    0 \\
  \end{array}
 \right)=
\left(
            \begin{array}{cc}
             -1 & 0 \\
              0 & 1 \\
            \end{array}
          \right)
\left(
  \begin{array}{c}
   -1 \\
    0 \\
  \end{array}
  \right)
  \end{equation}
which means that \begin{equation}\left(
  \begin{array}{c}
   1 \\
    0 \\
  \end{array}
  \right)
  =-\left(
  \begin{array}{c}
   -1 \\
    0 \\
  \end{array}
  \right).
  \end{equation}

It is clear that if the ket $\left|s\right>$ represents the spin
state and $U\in SU(2)$ then $\left<s|U^*U|s\right>$ is invariant
regardless of the representation. This is what we mean by $SU(2)$
spin invariance. We now separately analyze each representation.
Following the notation of Greenberger et al \cite{green}, we can
write $\left|\vec n,+ \right>$ and $\left|\vec n,- \right>$ to
represent spin-up and spin-down respectively along the $\vec n$
direction. Therefore,
\begin{equation}\left|\vec n_1,+\right>=(\cos
\theta/2)\left|\vec n_2,+\right> +(\sin \theta/2)\left|\vec
n_2,-\right>. \end{equation} Now choose $\vec n_2$ by rotating
through $\theta=\pi/2$ (clockwise) with respect to $\vec n_1$ and
choose $\vec n_3$ by rotating through $\theta=-\pi/2$
(anti-clockwise) with respect to $\vec n_1$ (Fig. 1). Then equation
(3) gives

\begin{eqnarray} \left|\vec n_1,+\right>&=&1/\sqrt 2\left|\vec n_2, +\right>+
1/\sqrt 2\left|\vec n_2,-\right>\\
&=&1/\sqrt 2\left|\vec n_3, +\right>- 1/\sqrt 2\left|\vec
n_3,-\right>.
\end{eqnarray}
\begin{pspicture}(-1,-2.5)(5,2.5)
\psline[linecolor=gray,linestyle=dashed, arrows=->](0,0)(4,0)
\rput(4.6,0){$\left|\vec n_1, +\right>$}
\psline[linecolor=red,linestyle=dashed, arrows=->](0,0)(2,2)
\rput(1,1.2){$\left|\vec n_2, +\right>$}
\psline[linecolor=green,linestyle=dashed, arrows=->](2,2)(4,0)
\rput(3,1.2){$\left|\vec n_2, -\right>$}
\psline[linecolor=green,linestyle=dashed, arrows=->](0,0)(2,-2)
\rput(1,-1.2){$\left|\vec n_3, +\right>$}
\psline[linecolor=red,linestyle=dashed, arrows=<-](2,-2)(4,0)
\rput(3,-1.2){$\left|\vec n_3, -\right>$}
\rput(8,0){{\small Fig. 1}}%
\end{pspicture}
In particular if $\left|\vec n_1,+\right> = \left(
                                              \begin{array}{c}
                                                1 \\
                                                0 \\
                                              \end{array}
                                            \right)$,
then $\left|\vec n_2,+\right>=\left(\begin{array}{c} 1/\sqrt
2\\1/\sqrt 2\\
\end{array}\right)$ and $\left|\vec n_2,-\right>=\left(\begin{array}{c} 1/\sqrt
2\\-1/\sqrt 2\\
\end{array}\right)$, and $\left|\vec n_3,+\right>=\left(\begin{array}{c} 1/\sqrt
2\\-1/\sqrt 2\\
\end{array}
\right)$ and
$\left|\vec n_3,-\right>=\left(\begin{array}{c}-1/\sqrt 2\\-1/\sqrt 2\\
\end{array}\right)$. Note, immediately that
$\left<\vec n_2,\pm|\vec n_3,\pm \right>=0$ and
$\left|n_2,-\right>=\left|n_3,+\right>$. This creates the ambivalent
situation of identifying, from the perspective of $\vec n_1$,  a
spin-up state ($\left|n_2,-\right>$) with a spin-down state
($\left|n_3,+\right>$), which forces the obvious question, as to
what is the meaning of up or down in this case. In fact from the
perspective of $\vec n_1$ alone there is no consistent meaning; for
the value depends not only on the direction $\vec n_1$ but on the
choice of angle $\theta$ defined relative to $\vec n_1$. It is
precisely this ambivalence that is at the heart of the GHZ
inconsistency and Bell's inequality.

\section {COPENHAGEN OR REALISM} In the light of the above relativistic
approach, we now ask how the above process can be viewed from both
the perspective of Bell's original paper on hidden variables
\cite{Bell} and from the perspective of the (weak) Copenhagen
interpretation. Bell assumes that ``if the two measurements are made
at places remote from one another the orientation of one magnet does
not influence the result obtained with the other"(II Formulation),
and suggests that if this ``locality" hypothesis fails then ``the
statistical predictions of quantum mechanics are incompatible with
separable, [local] predetermination" (V. Generalization). This calls
for some comments and observations.

(1) The word ``influence''in the first quote can have two meanings.
One meaning can refer to a physical ``influence" between the two
magnets, typified by some type of physical communication as in the
case of an electromagnetic transmission. The other ``influence'' can
come from a type of knowledge associated with the rules of
conditional probability and SU(2) relativity.

(2) In this paper, the term ``influence'' is only been used in the
second sense, which means we are assuming no physical contact
between the magnets and the Lorentz invariance is always maintained
with respect to space-time.

(3) The second type of influence allows for instantaneous knowledge
of physical events beyond the light cone, without in anyway
violating the laws of physics or causality. To understand this
better, let us consider the following analogous example. Alice and
Bob are each given a sealed envelope containing one of two possible
cards. Alice's envelope contains the Ace of Hearts, while Bob's
envelope contains the Ace of Spades. Each one of them is sent on a
long train journey going in opposite directions and are told to open
their envelopes after one hour. If Alice were asked, prior to
opening her envelope, the probability that Bob had the Ace of Hearts
she would answer 1/2. However, if she were asked the same questions,
after she had opened her envelope, she would have responded 0. In
other words, by Alice opening her first envelope, her new found
knowledge ``influences'' the probability outcome of Bob's
experiment. Knowledge, of the content of her envelope gives her
instantaneous knowledge of Bob's envelope. However, this
``influence'' is grounded in the rules of conditional probability
and the law of large numbers, without in any way violating the rules
of causality. Alice is not free to communicate her instantaneous
knowledge to Bob in an instantaneous way. Communication is subjected
to Lorentz invariance.

Similarly, in an analogous way, when one component of spin is
measured in a singlet state, we not only instantaneously know the
other but spin values can be preassigned by nature prior to the
experiment, although not necessarily in any deterministic way.
Moreover, precisely because of the SU(2) relativity effect (as
described previously), preassigned eigenstates (in a direction $\vec
n_1$) cannot be used to mathematically determine the preassigned
eigenstates of another direction, $\vec n_2$. Hence, our inability
to determine the values of preassigned states, is in full agreement
with Bell's assertion that ``the statistical predictions of quantum
mechanics are incompatible with separable predetermination'' of
observed values by means of a mathematical formula. However, it
would not be in agreement with the assertion that ``the statistical
predictions of quantum mechanics are incompatible with separable
predetermination'' in an ontological sense. In other words, nature
obeys the rules of causality but not necessarily in any systematic
way that precludes chance, nor in any absolute way to preclude
relativity by determining absolute values.

(4) Bell has rightly assumed that ``non-locality" implies ``no
separable predetermination'' of spin values by means of a
mathematical formula based on hidden parameters. However, he also
seems to imply a fallacy, when he concludes the opposite, namely
that a theory which involves ``no separable predetermination'' of
spin values could not be ``Lorentz invariant''. In contrast, this
paper has pointed out, that SU(2) relativity implies that there is
no separable predetermination of measurable spin values by
mathematical means but yet the theory is still local and hence
Lorentz invariant. This means that for any arbitrarily chosen
direction, the spin values are preassigned prior to the experiment.
However, precisely because of the SU(2) relativity effect (as
described above), preassigned eigenstates in one direction ($\vec
n_1$) cannot be used to determine preassigned eigenstates in another
($\vec n_2$). Nevertheless, cause and effect are preserved but not
systematically. The rules of chance always apply, and values cannot
be assigned to spin in an absolute way, for the same reason absolute
mass cannot be assigned to each of the particles. Finally, note that
the SU(2) frame of reference allows us to determine not the actual
outcome of the experiment but rather the probability distribution
for the spin-observables of the pre-correlated states.

(5) The Copenhagen interpretation will also undergo modifications,
depending on how one interprets the projection postulate. To better
understand the nature of the collapsed wave function and conditional
probability, we return to the probability experiment with Alice and
Bob. The initial state (for the Ace of Hearts and the Ace of Spades)
is given by
\begin{equation}
\left|\psi\right>=\left|A_h\right>\left|A_s\right>+\left|A_s\right>\left|A_h\right>.
\end{equation}
However, once either one of the envelopes are open, the state
changes (or collapses) for the person opening the envelope into
either $\left|\psi\right>= \left|A_h\right>\otimes \left|A_s\right>$
or $\left|\psi\right>= \left|A_s\right>\otimes \left|A_h\right>$
Note this is a consequence of conditional probability theory and a
natural explanation can be given in terms of moving from ignorance
to knowledge, without in any way suggesting that causality has been
violated. Similarly, in quantum mechanics an analogous situation
arises, although the property of rotational invariance and SU(2)
relativity involves purely a quantum phenomenon.

To summarize, from the perspective of this paper while the principle
of superposition as prescribed by the Copenhagen convention still
remains and the initial quantum state is best written as a complete
set of eigenvectors, it does not follow that the act of measuring
alone determines the observed eigenstate. Rather the measuring
devices because of its anisotropic nature, selects one of the
preexisting eigenstates as an axis of rotation, thus fixing the
preassigned information along that axis, while breaking the
preassigned correlated information in other directions because of
the rotation effect. Moreover, the anisotropic measurement increases
our knowledge of the situation in a given direction, while at the
same time destroying information about the initial isotropic state.
From a probability point of view, we can say that the measuring
device imposes a Markov condition on the newly emerging state by
disentangling the singlet. Moreover, once an observation is made,
the observed measurement can be identified with the collapsed wave
function in accordance with the mathematical rules for projection
operators.

\section {INEQUALITIES TO EQUALITIES}

In view of the above, we now re-formulate Bell's perspective from
the perspective of SU(2)relativity. In Bell's original paper, he
argues that for a system of particles in the singlet state the
expectation $E(\lambda_i^{(1)} \lambda_j^{(2)})$, identified with
``hidden'' parameters, is never equal to the quantum expectation
$\left<\sigma^{(1)}_i \sigma^{(2)}_j\right>$ where the $(1)$ and
$(2)$ refer to particle 1 and 2 respectively. We now proceed to
calculate this expected value, $E(\lambda_i^{(1)} \lambda_j^{(2)})$,
of the spin measured on a coupled system of particles, in the
directions $\vec n_i$ and $\vec n_j$. Recall $\vec n_i . \vec n_j =
\cos \theta_{ij} $. Let $P(\lambda_i^{(1)}, \lambda_j^{(2)})$ be the
joint distribution function associated with the two directions of
measurement then since the particles are in the singlet state
\begin{eqnarray}
E(\lambda_i^{(1)}\lambda_j^{(2)})&=&\sum \lambda_i^{(1)}
\lambda_j^{(2)}P(\lambda_i^{(1)} ,
\lambda_j^{(2)})\\
&=&\sum \lambda_i^{(1)} P_1(\lambda_i^{(1)})\lambda_j^{(2)}
P_{ji}(\lambda_j^{(2)}|\lambda_i^{(1)})\\
&=&(1)(\frac 12) (-1) \cos^2 (\frac {\theta_{ij}}{2})+(-1)(\frac
12) (1) \cos^2 (\frac {\theta_{ij}}{2})\nonumber\\
&\ &\qquad +(1)(\frac 12)(1)\sin^2 (\frac {\theta_{ij}}{2})
+(1)(\frac 12)(1)\sin^2 (\frac{\theta_{ij}}{2})\\
&=&-\cos^2 \frac {\theta_{ij}}{2} +\sin^2 \frac {\theta_{ij}}{2}\\
&=&-\cos \theta_{ij}\\
&=&-\vec n_i . \vec n_j\\
&=&\left<\sigma_i^{(1)} \sigma_j^{(2)}\right >.
\end{eqnarray}
Note our equation (7) is identical in form to equation (2) of Bell's
paper while our equation (12)and (13) are identical to equation (3)
of Bell's paper. In our case, however, both equations (7),(12) and
(13) coincide while in Bell's paper they do not. This difference
stems from the fact that Bell does not prioritize the direction of
the measuring magnetic field when dealing with hidden parameters and
his derivation presupposes that three independent measurements can
be made in three different and arbitrary directions without
consideration of the $SU(2)$ reference frame. Wigner's derivation of
a comparable inequality to that of Bell's rests on the same fallacy
\cite {wig}.

\section {CONCLUSION} In conclusion, note that the epistemological
model presented in this paper presents no difficulty in assigning a
real objective state to the elementary particle. However, a
numerical assignment of values to this objective state should
\newline (1) take into account the SU(2) relativity of the
situation,\newline (2) realize that such a state can be
measured,\newline (3) recognize that the measurement changes the
system by redefining the initial conditions (the uncertainty
principle),\newline (4) understand the limited way in which the
singlet state allows us to take a second measurement of the initial
system.

This last point takes us back to Einstein's original version of the
EPR paradox which was formulated in terms of position and momentum
and not spin \cite{Bell}. From the perspective of this paper,
Einstein was partially correct. Realism still exists in nature and
the projection postulate of QM, if viewed from the (weak) Copenhagen
perspective, violates this realism. It is unfortunate, therefore,
that the measurement of spin itself has become the standard for
testing the paradox. It seems to me that in reality the Alain Aspect
[1] experiment and other such experiments challenge the naive
realist interpretation of quantum mechanics, but they do not prove
Einstein's position to be incorrect. Naive realism is not realism,
it is just a flawed model of reality. SU(2) relativity not only
removes the inconsistency associated with naive realism by
preserving locality, but also seriously challenges the ``action at a
distance'' associated with the weak Copenhagen interpretation.
Indeed, from this perspective Einstein's objection is still valid.

On the other hand, Einstein's rejection of the statistical basis of
quantum mechanics stemming from the (weak) Copenhagen interpretation
cannot, in my opinion, be fully justified. A rejection of the
projection postulate cannot be used to reject a statistical
interpretation of quantum mechanics. In the above model, the quantum
state actually exists in an ontological way, as an element of
physical reality. It results from more than just our lack of
knowledge of initial conditions, as suggested by the ensemble
interpretation associated with Einstein. \footnote {Although, it
should be pointed out that the (weak) interpretation, is also
subjected to the law of large numbers and can also be viewed from
the ensemble perspective.}Indeed, the dynamical properties of spin
are not only affected by the measuring process, as exemplified in
the case of the spin-singlet state becoming disentangled, but they
do so in such a way as to maintain realism while avoiding the
difficulties associated with Bell's inequality.

Specifically, SU(2) relativity combined with conditional probability
theory and the projection postulate, not only guarantee realism but
also the necessity of a statistical interpretation of physical
reality, associated with Heisenberg's uncertainty principle. In
essence the uncertainty relations are a proof that a complete set of
initial conditions can never in principle be fully specified,
because the act of measurement itself changes the original system.
It is precisely this inability to specify precise initial conditions
that allows a probability theory to emerge. In this context, Lindsey
and Margenau note that the principle of indeterminacy in QM ``is
much more thoroughgoing, for it converts a question of convenience
into a matter of necessity. It expresses the conviction ultimately
that we shall never in any case be able to carry out even in
principle the measurement necessary for the exact determination of
boundary conditions in any physical problem''\cite{mar}. It then
follows from this that ``ignorance of them [boundary conditions]
forces us back on probability considerations''\cite{mar}.

Finally on a philosophical level we have modified the weak
Copenhagen interpretation of quantum mechanics with regards to its
interpretation of the projection postulate and its failure to grasp
SU(2) relativity. Our interpretation is based on a critical realist
approach which takes the given data to be objective, without falling
into some Newtonian dualistic interpretation of reality. We do not
put the observer outside of the world he is observing as is the
approach in classical mechanics or hidden variable theory (naive
realism being a case in point), nor do we go to the other extreme of
giving a singular importance to the observer as in the Copenhagen
interpretation. The measurement of spin is itself a SU(2) relativity
effect. We interact with the physical world because we are part of
the physical world and indeed it is precisely this interaction that
permits us to do physics. The laws of interaction are part of the
objective laws of the universe and are there to be discovered.
Quantum physics has tried to incorporate those interactions
pertaining to physics into its axioms. To the extent that it has
succeeded, quantum mechanics gives a more real and objective picture
of physical reality, the uncertainty relations being a case in
point. To the extent that these axioms are not yet complete (the
projection postulate), quantum mechanics should be willing to modify
its basic axioms in the light of the evidence.

\end{document}